\documentclass[12pt]{article}
\usepackage{amssymb,psfrag,amsmath,cite,bm}
\newtheorem{theorem}{Theorem}

\newtheorem{property}{Property}

\title{Optimal Control of Legged-Robots Subject to Friction Cone Constraints}

\author{Farhad Aghili\thanks{email:faghili@encs.concordia.ca}}

\begin{document}
\date{}
\maketitle

\begin{abstract}
A hierarchical control architecture is presented for energy-efficient control of legged robots subject to  variety of linear/nonlinear inequality constraints such as Coulomb friction cones, switching unilateral contacts, actuator saturation limits, and yet minimizing the power losses in the joint actuators. The control formulation can incorporate the nonlinear friction cone constraints into the control without recourse to the common linear approximation of the  constraints or introduction of slack variables. A performance metric is introduced that allows trading-off the multiple constraints when otherwise finding an optimal solution  is not feasible.  Moreover, the projection-based controller does not require the minimal-order dynamics model  and hence allows  switching contacts that is particularly appealing for legged robots. The fundamental properties of constrained inertia matrix derived are similar to those of general inertia matrix of the system and subsequently these properties are greatly exploited for control design purposes. The problem of task space control with minimum (point-wise) power dissipation subject to all physical constraints is transcribed into a quadratically constrained quadratic programming (QCQP) that can be solved by barrier methods. 
\end{abstract}

%------------------------------------------------------
\section{Introduction}
%------------------------------------------------------

Many robotic systems are subject to unilateral and friction cone constraints such as walking robotics
\cite{Aghili-2017a,Klein-Kittivatcharapong-1990,Gregorio-Ahmadi-Buehler-1997,Sangwan-Agrawal-2009,Braun-Mitchell-Goldfarb-2012,Saab-Ramos-Keith-2013,Prete-Mansard-Nori-2014,Park-Hong-Kim-2014,Farnioli-Gabiccini-Bicchi-2015,Caron-Pham-Nakamura-2015,Seok-Wang-Chuah-2015,Na-Kong-2015,Hunt-Giardina-2016},
cooperative robotic manipulators
\cite{Paljug-Yun-Kumar-1994,Aghili-Nmavar-2006,Aghili-Parsa-2009b}, and robotic grasping
\cite{Cole-Hauser-Sastry-1989,Bicchi-Kumar-2000,Saut-Remond-Perdereau-2005,Hirukawa-Hattori-Harada-2006,Zhu-Ding-2006,Boyd-Wegbreit-2007,Zhang-Jiang-Wu-2014,Zhang-Sun-Liu-2015,Zhang-Jiang-2015}.
All of these robotic  systems are subject to Coulomb friction cone constraints, which are essentially
nonlinear inequalities complicating the control
\cite{Aghili-2017a}. The control design can be even further complicated as some of these robotic systems are underactuated because of  their floating base, or they might have time-varying topology, where the total
degrees-of-freedom (DOF) exhibited  by the constrained robot changes
upon switching contact. In addition, the  number of actuators are usually higher
than the required number of DOF in the task space and hence there is redundancy. For practical purposes, it is highly desirable to exploit  the redundancy for minimizing  power losses in the joint motors. The advantages of doing so are twofold: i) alleviate overhearing of the motors, ii) increase the run-time of battery operated robots. From the preceding discussion, one can observe that these robotic systems ideally call for a single multi-objective
controller which can achieve task space tracking, accommodate switching contacts, handle linear and nonlinear inequality constraints pertinent to a variety of physical constraints, and yet can achieve energy-efficiency by minimizing
the power losses.

% Friction Cone-------------------------------
The prior arts  in control of robots subject to friction cone constrains
are essentially based on linearization of the nonlinear friction-cone
constraints for  linear programming implementations
\cite{Aghili-2017a,Kerr-Roth-1986,Cheng-Orin-1989,Caron-Pham-Nakamura-2015,Li-Ge-Ye-2015} or quadratic programming implementations \cite{Kuindersma-Permenter-2014,Righetti-Schaal-2012,Zhang-Zhang-2013,Escande-Mansard-Wieber-2014}.
Different linearization techniques have been introduced by
researchers to deal with the nonlinearity associated with the
Coulomb friction cones. Linearization of the nonlinear
friction-cone constraint for incorporation into a Linear
Programming (LP) for force optimization was first introduced in
\cite{Kerr-Roth-1986}. It was shown later that the linear
programming formulation for the optimal force control problem of a
walking robot with friction cone constraints can be simplified
through a proper use of pseudoinverse
\cite{Klein-Kittivatcharapong-1990}. Nonnegative slack variables
were introduced in order to approximate the nonlinear expression
of Coulomb friction cone by a four-sided friction pyramid
\cite{Yin-Hosoe-2004}. Just for a planar manipulation, this method
leads to eight linear inequalities in terms of four slack
variables. The contact stability of legged robots  under
the linear approximation of friction cones was presented in
\cite{Caron-Pham-Nakamura-2015}. Another approach to deal with the
friction cone constraint is to control the interaction forces in
such a way that the vector of contact force is oriented
vertically as much as possible by minimizing the
tangential force components and thereby  reducing the risk of
slipping during walking. An
optimization control method using the cost function similar to
\cite{Righetti-Buchli-Mistry-2011} is presented in
\cite{Prete-Mansard-Nori-2014} in order to enhance the flexibility of
optimization-based technique. However, it is known that the
linearization of the friction cone constant  is particularly
problematic for control because of the shortcomings associated with accuracy and stability of the linearization technique \cite{Kerr-Roth-1986,Borgstrom-Batalin-2010}. Moreover,
minimization of the tangential components of contact force is not
an optimal way to maintain the friction cone constraint because
this technique  may lead to excessive joint torques and subsequent power dissipations in the actuators.

In the realm of parallel robots,  the traditional inverse dynamic control method based on the independent coordinates \cite{Shang-Cong-Ge-2013,Aghili-Parsa-2007c} and the coordination motion control  \cite{Shang-Cong-2014} have been designed and implemented.
Nakamura et al. \cite{Nakamura-Ghodoussi-1989} derived dynamics equation of redundant or non-redundant parallel mechanisms based on coordinates partitioning for controls. The derivation of the dynamics formulation was simplified in \cite{Cheng-Yiu-Li-2003} to be used for implementation of the augmented PD control and computed-torque methods.
Traditional approaches for control of constrained robots are based on derivation of a reduced-order dynamics model in terms of an independent vector variables whose equals to the number of degrees-of-freedom exhibited by the constrained systems \cite{Raibert-Craig-1981,Khatib-1987,Yoshikawa-1987,Yoshikawa-Sugie-Tanaka-1988,Nakamura-Ghodoussi-1989,Doty-Melchiorri-1993,Luca-Manes-1994,Schutter-Bruyinckx-1996}.
These control approaches are  not adequate for robotic systems with time-varying topology because changing the number of degrees-of-freedom inevitably changes the number of independent variables that, in turn, requires switching to different dynamics models \cite{Aghili-2015b}. Alternatively, dynamics equation of constrained systems can be described in terms of dependent coordinates using linear projection framework so that the structures of dynamics model and the underlying control system remain unchanged even when the robotic system changes its number of
degrees-of-freedom \cite{Aghili-Piedboeuf-2003a,Aghili-Su-2016a,Aghili-Parsa-2007}. This control approach is proven to be particularly useful for control of legged robots with switching
contacts \cite{Aghili-2005,Righetti-Buchli-Mistry-2011}, or for
operational space control of a robot regardless of whether it  is constrained or unconstrained.

This paper calls for a projection-based multi-objective control
architecture for energy-efficient control of robots subject to a set
of linear and nonlinear inequality constraints, e.g., friction
cones constraints and actuator saturation limit. The controller is
general enough to accommodate changing constraint condition, e.g.,
switching between different contact modes, or to deal with
fully-actuated or underactuated robots alike \cite{Aghili-2017a}. It also allows  to
directly include nonlinear inequality constraints, e.g., friction
cone constraints, without recourse to linearization of constraints
or to introduce slack variables. Therefore, the controller can
robustly manage the interaction forces so that they lie within
their friction cones while achieving operational space control
with  minimum power dissipation in the actuators. In this
dynamic formulation, the fundamental properties of constrained
inertia matrix derived are similar to those of general inertia
matrix of the system making it particulary useful for control
design and analysis.  The optimization formulation also allows trading-off the space task with minimum perturbation to the trajectory tracking performance when it is not possible to enforce both physical limitations and operational space task requirement. This projection-based control approach is applicable for a range of robotic applications such as walking robots, grasping robots, as well as parallel robots. Section~\ref{sec:Modeling} reviews
non-reduced dynamics model of robots subject to a variety of
linear and nonlinear physical constraints.
Section~\ref{sec:ControlArchitecture} is devoted to present
systematic development of the control architecture. Operational
space control using  projection matrix is given in
Section~\ref{sec:OperationSpaceControl} followed by
Section~\ref{sec:OptimalControl} where the power loss cost
function along all linear and nonlinear physical constraints are
formulated as a quadratically constrained quadratic programming (QCQP) to be solved by barrier methods. Section~\ref{sec:tradeoff} presents a trade-off solution to the optimization problem when it is not possible to enforce all  physical limitations and the operational task requirement.

%-------------------------------------------------------------
\section{Non-reduced Dynamic Model Based on Oblique Projection}
\label{sec:Modeling}
%-------------------------------------------------------------
Suppose $\bm q\in\mathbb{R}^n$  be the generalized coordinates  of  a serial robotic arm whose open-loop dynamics  is characterized by inertia matrix $\bm M \in \mathbb{R}^{n \times n}$,  Coriolis term $\bm C(\bm q, \dot{\bm q}) \dot{\bm q} \in \mathbb{R}^{n}$, and gravitational force $\bm\tau_g$. Then, dynamics behavior of such robot interacting with the multiple frictional constraint surfaces can be described by
\begin{subequations} \label{eq:dynamics}
\begin{align} \label{eq:Mddq}
\bm M \ddot{\bm q} +  \bm C(\bm q, \dot{\bm q}) \dot{\bm q} & =
\bm B \bm u + \bm\tau_g - \bm A^T \bm\lambda \\
\label{eq:lamb_square} \text{subject to:} \qquad
\quad\sqrt{\lambda_{x_i}^2 +  \lambda_{y_i}^2} & < \mu_i
\lambda_{z_i} \\ \label{eq:lamb_z1} \lambda_{z_i} & > 0  \qquad
i=1,\cdots,k\\  \label{eq:actuator_limits} \bm u_{\rm min} &
\prec \bm u \prec \bm u_{\rm max},
\end{align}
\end{subequations}
where  $\bm A=[\bm A_1^T \; \bm A_2^T \; \cdots \; \bm A_k^T ]^T$ is the overall Jacobian matrix, $\bm A_i$ is the $3\times n$ translation Jacobian of the robot calculated at the $i$th contact point, $\bm B \in \mathbb{R}^{n \times p}$ is the actuator matrix owing to the $p$ actuated joints, $\bm u \in\mathbb{R}^p$ is the actuator force with the corresponding minimum and maximum limits  $\bm u_{\rm min}$ and $\bm u_{\rm max}$, and $\bm\lambda =[\bm\lambda_1^T \; \bm\lambda_2^T \; \cdots \; \bm\lambda_k^T ]^T\in \mathbb{R}^m$ is the  generalized lagrangian multipliers and $\bm\lambda_i=[\lambda_{x_i}\; \lambda_{y_i} \; \lambda_{z_i}]^T$ constitutes the $3\times1$ force vector at the $i$-th contact point. The lagrangian multiplier vector at frictional contact surface $i$ with the coefficient of friction $\mu_i$  is defined in such a way that $\lambda_{z_i}$ is normal to the contact surface while  $\lambda_{x_i}$ and $\lambda_{y_i}$ become tangential to the constraint surface. Thus, inequality constraints \eqref{eq:lamb_square} and \eqref{eq:lamb_z1} are enforced to retain  the interaction forces to lie within their friction cones and to ensure they  always  push and not pull on at the contact surfaces, while inequalities \eqref{eq:actuator_limits} ensure the actuator forces do not exceed their saturation limits. It is worth noting that since the sliding friction is avoided by enforcing the friction cone constraints,  the sliding velocity between the contact surfaces is zero.

Under the $k$ contact constraints, the constraint equations at the velocity level are described by
\begin{equation} \label{eq:Adotq}
\bm A(\bm q) \dot{\bm q} = \bm 0 \in \mathbb{R}^m,
\end{equation}
where the Jacobian matrix $\bm A\in\mathbb{R}^{m\times n}$  may not be full-rank, that is $\text{rank}(\bm A)=r$ and $r\leq m$. Alternatively, the constraint condition \eqref{eq:Adotq} can be equivalently described by  $\bm P \dot{\bm q}= \dot{\bm q}$ \cite{Aghili-Piedboeuf-2003a}, where the square matrix $\bm P \in \mathbb{R}^{n \times n}$ represents the {\em orthogonal projection}\footnote{The projection matrix
can be obtained from $\bm P \triangleq \bm I -  \bm A^+ \bm A$ where $\bm A^+$ is the pseudo-inverse of $\bm A$, and $\mbox{rank}({\bm P}) = \mbox{tr}({\bm P}) =n-r$. Then, $\bm P^2= \bm P = \bm P^T$} onto the null-space of $\bm A$. Knowing that $\bm P \bm A^T = \bm 0$, one can eliminate the lagrangian multipliers by pre-multiplying  both sides of \eqref{eq:Mddq} by $\bm P$
\begin{equation} \label{eq:projected_dynamics}
\bm P \bm M \ddot{\bm q} + \bm P \bm C \dot{\bm q}  =
 \bm P(\bm B \bm u + \bm\tau_g),
\end{equation}
As shown in Appendix~\ref{appx_skew}, one can also show that the generalized velocity is linearly mapped to  the ${\cal N}^{\perp}$ component of the generalized acceleration $\ddot{\bm q}_{\perp} := ({\bm I - \bm P}) \ddot{\bm q}$ by
\begin{equation}\label{eq:Omega_dq}
\ddot{\bm q}_{\perp} = \bm\Omega \dot{\bm  q},
\end{equation}
where $\bm\Omega =\bm L^T - \bm L $ is a skew-symmetric, and $\bm L = - \bm A^+ \dot{\bm  A} \bm P$.

As elaborated in Appendix~\ref{appx:formulation}, the kinematics equation \eqref{eq:Omega_dq} can be combined with the dynamics equation \eqref{eq:projected_dynamics} to obtain  the
following non-minimal order dynamics model
\begin{subequations} \label{eq:constraint_system}
\begin{equation} \label{eq_EqMotion2}
\bar{\bm M}(\bm q) \ddot{\bm q} + \bar{\bm C}(\bm q, \dot{\bm q})
\dot{\bm q}   = \bm P \bm B \bm u + \bm P \bm\tau_g,
\end{equation}
where
\begin{align} \label{eq:Mbar}
\bar {\bm M} (\bm q) & =   \bm P    \bm M(\bm q)   \bm P + \nu
(\bm I -\bm P),\\ \label{eq:Cbar2} \bar{\bm C}(\bm q, \dot{\bm q})
& = \bm P \bm C(\bm q, \dot{\bm q}) \bm P + \bm P \bm M (\bm\Gamma
+ \bm\Gamma^T) - \nu \bm\Gamma,
\end{align}
\end{subequations}
and $\nu>0$ is any positive scalar to define the  constraint inertia matrix $\bar{\bm M}$.
\begin{property} \label{eq:prop_constraint_mass}
The constraint inertia matrix has the following fundamental
properties:
\begin{enumerate}
\item $\bar{\bm M}$ is always  positive-definite, bounded, and symmetric
matrix. \item $\frac{d}{dt}{\bar{\bm M}} - 2 \bar{\bm C}$ is a
skew-symmetric matrix.
\end{enumerate}
The proof is given in Appendix~\ref{appx:formulation}.
\end{property}
%From the numerical efficiency point of view, it should be pointed out that the condition number of the %constraint inertia matrix can be minimized if the $\nu$ is selected to be within a specific range  \cite{Aghili-2015b}.
Finally, upon substituting the generalized acceleration obtained from
\eqref{eq_EqMotion2} into the original dynamics formulation
\eqref{eq:Mddq}, we arrive at the expression of the constraint
force as follow
\begin{equation} \label{eq:lambda}
\bm A^{T} \bm\lambda = -\bm S \big( \bm B \bm u + \bm\tau_g - \bm Q
\dot{\bm q} \big)
\end{equation}
where $\bm Q = \bm M \bm\Omega + \bm C$, and
\begin{equation}  \label{eq:S_definition}
\bm S :=\bm I- \bm M \bar{\bm M}^{-1} \bm P
\end{equation}
is an oblique projection matrix, i.e., $\bm S^2=\bm S \neq \bm
S^T$, see Appendix~\ref{appx:lagrangian} for details.

%Fig.1 =======================================================
%\begin{figure}
%\psfrag{N(A)}[c][c][.7]{${\cal N}({\bm A})$}
%\psfrag{NAperp}[c][c][.7]{${\cal N}^{\perp}({\bm A})$}
%\psfrag{R(LT)}[c][c][.7]{${\cal R}({\bm\Lambda}^T)$}
%\psfrag{R(B)}[c][c][.7]{${\cal R}(\bm B)$}
%\psfrag{Rn}[c][c][.7]{$\mathbb{R}^n$}
%\centering{\includegraphics[clip,width=8.5cm]{subspace}}
%\caption{The control subspaces: ${\cal
%R}(\bm\Lambda^T) \subset {\cal N}(\bm A) \cap {\cal R}(\bm B)$ (left);
%${\cal R}(\bm\Lambda^T) \equiv {\cal N}(\bm A)$ (right).}
%\label{fig:set_diagram}
%\end{figure}
%=============================================================

%-------------------------------------------------------------
\section{Control Architecture}
\label{sec:ControlArchitecture}
%-------------------------------------------------------------
The $r$  independent constraints means that the global DOF exhibited by the constrained system is $l=n-r$. Thus, it should be possible, in principle, to control a desired variable $\bm x(\bm q)~\in~\mathbb{R}^l$. Using the chain rule, the time-derivative of control variable yields
\begin{equation} \label{eq_dtheta}
\dot{\bm x}= \bm\Lambda \dot{\bm q}, \quad \text{where}\quad
\bm\Lambda \triangleq  \frac{\partial \bm x(\bm q)}{\partial \bm
q} \bm P.
\end{equation}
The acceleration in the operational space  can be obtained from
differentiation of \eqref{eq_dtheta} as
\begin{equation} \label{eq:ddx}
\ddot{\bm x}= \dot{\bm\Lambda} \dot{\bm q} + \bm\Lambda \ddot{\bm
q}
\end{equation}
%In following analysis, we shall  derive a minimal subspace of the generalized force required to control the %operational task space. This subspace reduction technique offers the possibility of greater extension of the remaining  %input-force utilization  for satisfying the physical constraints or/and for minimizing the cost function.

It can be inferred from \eqref{eq_dtheta} that ${\cal
N}(\bm\Lambda)$ component of the generalized velocity does not
generate any motion at the operational space meaning that
only the ${\cal N}^{\perp}(\bm\Lambda)={\cal
R}(\bm\Lambda^T)$ component of $\dot{\bm q}$ is required for the
desired motion generation. On the other hand, the kinematics constraint
specifies that any admissible $\dot{\bm q}$ should also belong to
${\cal N}(\bm A)$. Therefore, the following set
relationship is in order
\begin{equation} \label{eq:LTsubA}
{\cal R}(\bm\Lambda^T) \subseteq {\cal N}(\bm A)
\end{equation}
The null-space completely encapsulates the entire designated operational space if ${\cal R}(\bm\Lambda^T)={\cal N}(\bm A)$. Moreover, if the the control variables $\dot{\bm x}$ are selected to be a complete set of independent generalized velocities, then matrix $\bm\Lambda$ has linearly independent rows and hence it is full-rank. In this case, matrix $\bm\Lambda$ satisfies the following useful properties:
\begin{subequations}
\begin{align} \label{eq:PLambdaT}
\bm P \bm\Lambda^T &= \bm\Lambda^T \\ \label{eq:PLambda+}
(\bm I - \bm P) \bm\Lambda^+ & = \bm 0\\ \label{eq:Lambda+ddx}
\ddot{\bm q} - \bm\Gamma \dot{\bm q} & = \bm \Lambda^+ \ddot{\bm x}
\end{align}
\end{subequations}
where $\bm\Gamma = \bm\Lambda^+ \dot{\bm\Lambda} - \bm\Omega$; see Appendix~\ref{appx:Lambda} for details.

Let us define the vector of generalized control input $\bm\tau_c\in {\cal N}(\bm A)$. Then, if $\bm\tau_c$ and $\bm u$ are treated respectively as known and unknown variables, we are interested for find $\bm u$ so that the following constraint  is in order
\begin{equation} \label{eq:RBtau_parallel}
\bm\phi(\bm u)= \bm 0, \quad \mbox{where} \quad  \bm\phi(\bm u) \equiv \bm\tau_c -  {\bm P} \bm B \bm u.
\end{equation}
Since $\bm\tau_c \in {\cal N}(\bm A)$, the condition for existence of at least one solution for $\bm u$ is that ${\cal R}({\bm P}) \subseteq {\cal R}( {\bm P} \bm B)$. This means that $\bm B$ should not reduce the range of
${\bm P}$ and that is possible only if ${\cal R}({\bm
P}) \subseteq {\cal R}(\bm B)$, or equivalently
\begin{equation} \label{eq:Range-PB}
{\cal R}(\bm\Lambda^T) \subseteq {\cal R}(\bm B)
\end{equation}
In other words, if condition \eqref{eq:Range-PB} is met, then the
reciprocal of relation \eqref{eq:RBtau_parallel} exists and the minimum norm
solution can be described by pseudo-inversion
\begin{equation} \label{eq:mintau_a}
\bm u =  (\bm P \bm B)^+ \bm\tau_c
\quad \leftarrow \quad \min \| \bm u \|
\end{equation}
Both set relationships \eqref{eq:LTsubA} and \eqref{eq:Range-PB} come down to the following relationship
\begin{equation} \label{eq:LsubA}
{\cal R}(\bm\Lambda^T) \subseteq {\cal N}(\bm A) \cap {\cal R}(\bm
B),
\end{equation}
In other words, \eqref{eq:LsubA} implies: i)  Consistency of the specified operational space variable $\bm x(\bm q)$ with respect to  the admissible null-space
motion; ii) feasibility of the task space control given the structure of the actuator matrix $\bm B$.

\subsection{Projection-Based Operation Space Control}
\label{sec:OperationSpaceControl}

\subsubsection{Reference tracking}
Let us rearrange  \eqref{eq:projected_dynamics} in the following form
\begin{equation} \label{eq:PxMddq}
{\bm P} \bm B \bm u ={\bm P} \bm M \ddot{\bm q} + {\bm P} \bm C \dot{\bm q} -   {\bm P} \bm\tau_g.
\end{equation}

\begin{theorem} \label{theorem:PIDC}
Consider  independent set $\bm x(\bm q) \in \mathbb{R}^l$ for the operational space control. Suppose
that condition \eqref{eq:LsubA} holds and hence there exists $\bm u$ satisfying  constraint \eqref{eq:RBtau_parallel}, where ${\bm\tau}_c$ is given by the following expression
\begin{equation} \label{eq:PIDC}
{\bm\tau}_c  = {\bm P} \bm C \dot{\bm q} -{\bm P}
\bm\tau_g + {\bm P} \bm M \big(\bm\Lambda^+
(\ddot{\bm x}_d + \bm K_D \dot{\bm e} + \bm K_P \bm e)
- \bm\Gamma \dot{\bm q} \big),
\end{equation}
$\bm e=\bm x_d - \bm x$, $\bm x_d$ is the reference position trajectory, and
$\bm K_P \succ 0$ and $\bm K_D \succ 0$ are positive gain
matrices.  Then $\bm x(t) \rightarrow \bm x_d(t)$ as $t \rightarrow \infty$.
\end{theorem}
{\sc Proof:}  In view of \eqref{eq:RBtau_parallel}, and upon  substitution of the generalized force
from \eqref{eq:PIDC} into \eqref{eq:PxMddq}, we can derive the dynamics equation of the closed-loop system
\begin{align} \notag
{\bm P} \bm M \ddot{\bm q} &+ {\bm P} \bm C \dot{\bm
q} - {\bm P} \bm\tau_g =  \bm P \bm C \dot{\bm q} - {\bm P} \bm\tau_g
\\ \label{eq:error_dyanmics} &\qquad +  \bm P \bm M \Big( \bm\Lambda^+( \ddot{\bm x}_d
+ \bm K_D \dot{\bm e} + \bm K_P \bm e) -
 \bm\Gamma\dot{\bm q} \Big).
\end{align}
After simplification and upon substituting the expression of the acceleration from \eqref{eq:Lambda+ddx},   equation \eqref{eq:error_dyanmics} can be concisely written as
\begin{equation} \label{eq:diffe}
{\bm P} \bm M \bm\xi = \bm 0
\end{equation}
where $\bm\xi = \bm\Lambda^+ \bm\zeta$ and
\begin{equation} \label{eq:zeta_e}
\bm\zeta = \ddot{\bm e} + \bm K_D \dot{\bm e} + \bm K_P {\bm e}
\end{equation}
Identity  \eqref{eq:PLambda+} implies
$(\bm I - {\bm P}) \bm\xi = \bm 0$. Therefore, analogous to \eqref{eq:Mbar},
one can combine the latter identity and \eqref{eq:diffe} to derive the error dynamics equation in terms of the constraint inertia matrix by
\begin{equation} \label{eq:Mbar_xi}
\bar{\bm M} \bm\xi = \bm 0
\end{equation}
On the other hand, the only possibility for \eqref{eq:Mbar_xi} to
happen is that $\bm\xi \equiv \bm 0$ because $\bar{\bm M} \succeq 0$, and that in turn implies  $\bm\zeta
=0$ because $\bm\Lambda^+$ is a full-rank matrix. Therefore, one can conclude from \eqref{eq:zeta_e} that the position error exponentially converges to zero.  Subsequently, the minimum-norm
actuation force to produced the torque control law can be obtained be substituting $\bm\tau_c$ from \eqref{eq:PIDC} into \eqref{eq:mintau_a}.

Although  equality constraint \eqref{eq:RBtau_parallel} ascertains
operational space control of the constrained robot, the physical
constraints \eqref{eq:lamb_square}-\eqref{eq:actuator_limits} are
yet to be accounted for by the  hierarchical control architecture.
Section~\ref{sec:OptimalControl} will address point-wise optimal
control of the robotic system along its desired trajectories.

\subsubsection{Regulation}

Suppose the objective is to regulate independent variable $\bm x(\bm q) \in \mathbb{R}^l$ according to set point $\bm x_d$. Then, consider the following torque control law
\begin{equation} \label{eq:regulation}
\bm\tau_c  = \bm P \big( - \bm\tau_g
- \bm K_D \dot{\bm q} + \bm\Lambda^T \bm K_P \bm e \big),
\end{equation}
where $\bm e = \bm x_d - \bm x$. Using  \eqref{eq:PLambdaT} and  substituting $\bm\tau_c= \bm P \bm B \bm u$ from \eqref{eq:regulation} into \eqref{eq_EqMotion2}, we arrive at closed-loop dynamics of the control system as
\begin{equation} \label{eq:error}
\bar{\bm M} \ddot{\bm q} + \bar{\bm C} \dot{\bm q} = - \bm P \bm K_D
\dot{\bm q} +  \bm\Lambda^T \bm K_P \bm e,
\end{equation}
in which we use identity $\bm P \bm\Lambda^T = \bm\Lambda^T$ from \eqref{eq:PLambda+}. Define Lyapunov function
\begin{equation} \notag
V= \frac{1}{2} \dot{\bm q}^T \bar{\bm M} \dot{\bm q} + \frac{1}{2}
\bm e^T \bm K_P \bm e.
\end{equation}
By virtue of Property~\ref{eq:prop_constraint_mass}, the  time-derivative of the function becomes
\begin{align*}
\dot{V} &= \dot{\bm q}^T \big( \frac{1}{2} \dot{\bar{\bm M} }
- \bar{\bm C} \big) \dot{\bm q} - \dot{\bm q}^T \bm P \bm K_D \dot{\bm q} + \dot{\bm q}^T \bm\Lambda^T \bm K_p \bm e \\
&  - \dot{\bm q}^T \bm\Lambda^T \bm K_p \bm e  = -\dot{\bm q}^T
\bm K_D \dot{\bm q} < 0,
\end{align*}
which is negative-semidefinite. Clearly, we have $\dot V=0$ only if
$\dot{\bm q}= \bm 0$. Now, by substituting  $\ddot{\bm q}=\dot{\bm q}=
\bm 0$ in~\eqref{eq:error}, we can find the largest invariant set
with respect to system ~\eqref{eq:error} as the following
\begin{equation} \label{eq:Omega}
{\cal I} = \{\bm e, \dot{\bm q}: \; \dot{\bm q}= \bm 0 \quad \wedge \quad
\bm\Lambda^T \bm K_p \bm e= \bm 0 \}
\end{equation}
Since both $\bm\Lambda^T$ and $\bm K_p$ are full-rank matrices, the only possibility for $\bm\Lambda^T \bm K_p \bm e= \bm 0$ to occur is that $\bm e = \bm 0$. In other words, ${\cal I}=\{
\dot{\bm q} = \bm 0 \quad \wedge \quad \bm e=\bm 0\}$ is the largest invariant set which satisfies $V=0$. Therefore, according to LaSalle's Global Invariant
Set Theorem \cite{Lasalle-1960}, \cite[p. 115]{Khalil-1992}, the
solution of system~\eqref{eq:error} asymptotically converges to
the invariant set ${\cal I}$. Consequently, as the time progresses,
$\bm x$ asymptotically approaches its desired value $\bm
x_d$. The minimum actuator torque can be obtained by substituting  \eqref{eq:regulation} into \eqref{eq:mintau_a}.

\subsection{Optimization of control input} \label{sec:OptimalControl}
In this section, we complement the torque control law in order  to fulfill secondary control
objectives pertinent to the cost function  and the physical constraints.

\subsubsection{Cost function}
In the case that there are more actuators than the
degrees-of-freedom specified in the operational space, there is
redundancy meaning that there are infinite number of
solution to the inverse dynamics problem \cite{Xia-Feng-Wang-2005,Shang-Cong-Ge-2012}. The redundancy can be
exploited to minimize power losses in electric actuators by proper use of a pseudo-inverse technique.
Although solution \eqref{eq:RBtau_parallel} minimizes the Euclidean norm of the
actuator, the latter cost function is not with a clear physical
sense. In this section, we show that the redundancy can be
exploited to minimize power losses in electric actuators.
It should be pointed out that energy efficiency is highly desirable for autonomous robotic
systems carrying their own battery packs. This is because less
power dissipation means less heat generation in the robot actuators that
can avert the overheating problem. Furthermore, energy efficiency control of robot allows to use a smaller battery pack. The power losses in electromechanical system is indeed tantamount to the weighted
Euclidean norm of actuation force if the weighting matrix is
specifically selected \cite{Aghili-2011m}. According to the Ohms law, the collective
power losses in the joint motors is
\begin{equation} \label{eq:Ploss}
P_{\rm loss}= \bm i^T \bm R \bm i,
\end{equation}
where vector  $\bm i$ contains the motor currents and diagonal matrix $\bm R$ contains  winding resistance of the motors. On the other hand, since the torque of $j$th
motor is related to its current through the torque constant, i.e.,
$u_j = K_{t_j} i_j$, the quadratic function \eqref{eq:Ploss} can be equivalently
written as  \cite{Aghili-Buehler-Hollerbach-2000b}
\begin{equation} \label{eq:weighted_norm}
P_{\rm loss} = \| \bm u\|_W^2 = \bm u^T \bm W \bm u,
\end{equation}
where the $j$th diagonal entry of  the weighting matrix $\bm W$ is given by $W_{jj}=R_j/ K_{t_j}^2$.

\subsubsection{Physical Constraints}
Let us define selection vectors $\bm\sigma_{x_i}$, $\bm\sigma_{y_i}$, and $\bm\sigma_{z_i}$ in order to extract $x$, $y$, and $z$ components of the $i$-th contact according to $\lambda_{x_i} = \bm\sigma_{x_i}^T \bm\lambda$, $\lambda_{y_i} = \bm\sigma_{y_i}^T \bm\lambda$, and $\lambda_{z_i} = \bm\sigma_{z_i}^T \bm\lambda$. In the case of deterministic  lagrangian multipliers, the normal and tangential components of the $i$-th contact force  can be derived from the expression of the generalized lagrangian multipliers \eqref{eq:lambda} as the following
\begin{equation} \label{eq:lambda_xyz}
\begin{split}
\lambda_{x_i} &= \bm a_{x_i}^T \bm S(\bm B \bm u + \bm\tau_g - \bm Q \dot{\bm q}) \\
\lambda_{y_i} &= \bm a_{y_i}^T \bm S(\bm B \bm u + \bm\tau_g - \bm Q \dot{\bm q}) \\
\lambda_{z_i} &= \bm a_{z_i}^T \bm S(\bm B \bm u + \bm\tau_g -
\bm Q \dot{\bm q}) \qquad i=1,\cdots,k
\end{split}
\end{equation}
where $\bm a_{x_i} = \bm A^{+} \bm\sigma_{x_i}$, $\bm a_{y_i} = \bm A^{+} \bm\sigma_{y_i}$, and $\bm a_{z_i} = \bm A^{+} \bm\sigma_{z_i}$. To this end, substituting \eqref{eq:lambda_xyz} into the inequalities \eqref{eq:lamb_square} and \eqref{eq:lamb_z1}, the latter inequity constraints can be equivalently transcribed   by
\begin{subequations} \label{eq:zu_uGu}
\begin{align}
\bm z_i^T \bm u + \alpha_i & \geq  0 \\
\bm u^T \bm G_i  \bm u + \bm\gamma_i^T \bm u +\beta_i
& \geq 0,
\end{align}
\end{subequations}
where
\begin{equation} \notag
\begin{split}
\bm z_i & = \bm B^T \bm S^T \bm a_{z_i}\\
\bm\Pi_i &= \bm S^T \Big(-\bm a_{x_i} \bm a_{x_i}^T - \bm a_{y_i}
\bm a_{y_i}^T + \mu_i^2\bm a_{z_i} \bm a_{z_i}^T \Big) \bm S\\
\bm G_i & = \bm B^T \bm\Pi_i \bm B\\
\alpha_i &= \bm a_{z_i}^T \bm S  (\bm Q \dot{\bm q}- \bm\tau_g) \\
\beta_i  &= (\bm Q \dot{\bm q}- \bm\tau_g)^T   \bm\Pi_i  (\bm Q \dot{\bm q} - \bm\tau_g) \\
\bm\gamma_i & = 2 \bm B^T \bm\Pi_i (\bm\tau_g- \bm Q \dot{\bm q})
\end{split}
\end{equation}

\subsubsection{Quadratically constrained quadratic programming}
In view of  \eqref{eq:actuator_limits}, \eqref{eq:weighted_norm}, \eqref{eq:zu_uGu}, and \eqref{eq:RBtau_parallel}, the problem of the optimal control $\bm u$ minimizing power dissipation while satisfying the physical constraints is transcribed by  {\em
quadratically constrained quadratic programming} (QCQP)
\begin{subequations} \label{eq:quadratic_programming}
\begin{align} \label{eq:min}
\text{minimize} \qquad &  P_{\rm loss}(\bm u) =  \| \bm u \|_W^2 \quad &  \\
\label{eq:opt_equality} \text{subject to} \qquad & \bm\phi(\bm u) =\bm\tau_c - {\bm P} \bm B
\bm u   = \bm 0 \\ \label{eq:inequalities}
& \bm c(\bm u) \succeq   \bm 0,
\end{align}
where vector
\begin{equation} \label{eq:c_constraints}
\bm c(\bm u) = \begin{bmatrix}
\bm z_1^T  \bm u +
\alpha_1 \\
\bm u^T \bm G_1 \bm u + \bm\gamma_1^T \bm u +
\beta_1 \\
\vdots \\
\bm z_k^T  \bm u +
\alpha_k \\
\bm u^T \bm G_k \bm u + \bm\gamma_k^T \bm u +
\beta_k \\
-\bm u + \bm u_{\rm max} \\
\bm u - \bm u_{\rm min}
\end{bmatrix} \in \mathbb{R}^r
\end{equation}
\end{subequations}
contains a set of $r=2(k+p)$ linear and quadratic functions. Recall that the equality constraint \eqref{eq:opt_equality}  ensures operational
space control, the quadratic inequalities \eqref{eq:inequalities} are related to physical constraints due to  unilateral contacts, friction cones, and actuator
saturation limits, while the cost function \eqref{eq:min} represents total power loss in the
joint actuators.

All inequality constraints in \eqref{eq:quadratic_programming} can be reformulated as {\em log-barrier function} in the cost function of the following optimization programming, which has only equality constraint,
\begin{subequations} \label{barrier_problem}
\begin{align} \label{eq:min_log}
\text{minimize} \qquad &  J= \| \bm u \|_W^2 - \eta \sum_{i=1}^{r} \log c_i(\bm u)  \quad &  \\
\label{eq:constraint_log} \text{subject to} \qquad  &  \bm\phi(\bm u )  = \bm 0,
\end{align}
\end{subequations}
and $\eta > 0$ is the barrier parameter. As $\eta$ converges to zero, the solutions of \eqref{barrier_problem} should converge to that of \eqref{eq:quadratic_programming}. Thus, the original problem  is converted into the  smooth convex problem to be solved by a numerical technique such as the the Newton's method. Now, by defining the lagrangian function ${\cal L}= J + \bm\omega^T \bm\phi$ with  $\bm\omega$ being the dual variable associated with the equality constraint \eqref{eq:constraint_log}, the Karush-Kuhn-Tucker conditions stipulate that the optimal solution $\bm u^*$ should satisfies: i) $\bm\phi(\bm u^*)= \bm 0$, and ii) the stationary condition:
\begin{align*}
&\frac{\partial }{\partial \bm u}\big(  \| \bm u \|_W^2  - \eta \sum_i \log \psi_i(\bm u) + \bm\phi^T \bm\omega \big) \Big|_{\bm u^*} = \\
&  2  \bm W \bm u^* + \bm B^T {\bm P} \  \bm\omega  -  \eta \bm D^{-1}  (2 \bm u^* - \bm u_{\rm min} - \bm u_{\rm max} )\\
& - \eta \sum_{i=1}^k  \big( \frac{2 \bm G_i \bm u^* +
\bm\gamma_i}{\bm u^{*T} \bm G_i \bm u^* + \bm\gamma_i^T
\bm u^* + \beta_i} + \frac{\bm z_i}{\bm z_i^T \bm u^* -
\alpha_i} \big)   = \bm 0
\end{align*}
where $\bm D$ is a diagonal matrix whose diagonal entries are  $D_{ii}= u_i^{*2}  -
(u_{{\rm max}_i} + u_{{\rm min}_i}) u^{*}_i + u_{{\rm max}_i} \cdot u_{{\rm min}_i} $. Thus, we have now sufficient equations to obtain the solution of both $\bm\omega$ and $\bm u^*$ variables for a given barrier parameter $\eta$, and the solution converges to that of the original problem \eqref{eq:quadratic_programming} as the barrier parameter approaches to zero. However, assigning very small value for $\eta$ at the beginning  may lead to instability of the numerical solver and hence it should be gradually decreased, e.g.,  following the central path by using interior point methods \cite{Boyd-Vandenberghe-2004}; see Appendix~ \ref{appx:interior_point}. 

It is worth mentioning that although the robot is controlled in the operational space space $\bm x(\bm q)$ to perform some desired basic task, the redundancy can be also utilized by imposing a set of constraints on the dependent coordinates $\bm q$  to accomplish appropriate additional tasks such as avoiding singularity or joint limits \cite{Seraji-1989,Chaumette-Marchand-2001,Han-Park-2013}. The  additional task in the joint space cab important for robotic applications where the  workspace is limited by singularities and
joint limits \cite{Canudas-Siciliano-Bastin-book-1996,Chaumette-Marchand-2001,Han-Park-2013}. There are a variety of methods in the literature for dealing with
singularities and joint limits using  hierarchically structured
controller in the operational space control framework. For example there exists the classical gradient projection approach while the other more efficient method is based on activation function that does not affect the task achievement and ensures the avoidance problem \cite{Chaumette-Marchand-2001}. The basic idea is to added a feedback component to the task space controller by defining a function of $\bm q$ and their lower and upper limits  so that the additional feedback is activated only when  a joint angle lies outside its range and deactivated otherwise.

\subsection{Trade-off between multiple constraints} \label{sec:tradeoff}
Clearly, the necessary condition for existence of an optimal solution to \eqref{eq:quadratic_programming} is that both equality and inequality constraints \eqref{eq:opt_equality} and \eqref{eq:inequalities} are feasible. However, it may not possible to enforce both sets of constraints and hance a  tread-off between the constraints may become necessary. The inequality constraints can not be traded-off because they enforce  physical limitations of the system  pertaining to friction cone, unilateral contact, and actuator torque limits. On the other hand, the equality constraint enforces the operational space task and therefore  relaxing the equality constraint \eqref{eq:opt_equality} will trade-off the space task performance.  In this section we show that relaxing the constraint associated with the operational space task amounts to a disturbance vector entering to the position tracking system. Subsequently, the Euclidean norm of the disturbance is selected  as a performance metric to be minimized by adding that to the cost function. If the  the control constant  \eqref{eq:RBtau_parallel}   is no longer zero, i.e., $\bm\phi \neq \bm 0$, then the actuator torques can not generate the torque control law dictated by the motion controller \eqref{eq:PIDC}. In that case, we are interested to find out the impact of  non-zero residual $\bm\phi \neq \bm 0$ to the tracking performance.
By virtue of \eqref{eq:RBtau_parallel}, we can subtract both sides of equations \eqref{eq:PxMddq} and \eqref{eq:PIDC} to arrive at
\begin{align} \notag
\bm\phi &= \bm P \bm M \big( \bm\Lambda^+(\ddot{\bm x}_d + \bm K_D \dot{\bm e} + \bm K_D \bm e)  - \bm\Gamma \dot{\bm q} \big)   - \bm P \bm M \ddot{\bm q} \\ \notag
&= \bm P \bm M \bm\Lambda^+ \bm\zeta \\ \notag
&= (\bm P \bm M + \nu (\bm I - \bm P)) \bm\Lambda^+ \bm\zeta = \bar{\bm M} \bm\Lambda^+ \bm\zeta
\end{align}
where variable $\bm\zeta$ is previously defined by the second order differential equation \eqref{eq:zeta_e}; note that $(\bm I - \bm P)) \bm\Lambda^+=\bm 0$ because of \eqref{eq:PLambda+}.
Since $\bar{\bm M}$ is invertible, the above equation can be solved for vector $\bm\Lambda^+ \bm\zeta$ using matrix inversion
\begin{equation} \label{eq:e_dist}
\bm\Lambda^+(\ddot{\bm e} + \bm K_D \dot{\bm e} + \bm K_P \bm e) = \bm d, \quad \mbox{where} \quad \bm d= \bar{\bm M}^{-1} \bm\phi
\end{equation}
Clearly the error dynamics is represented by a non-homogeneous differential equation and  vector $\bm d$
enters as the disturbance. It is logical to minimize the magnitude of the disturbance in order to minimize the side effect of not fully satisfying the equality constraint as $\bm\phi \neq \bm 0$. Therefore, we consider the following mixed cost function
\begin{equation} \notag
\| \bm u \|_W^2 + \rho \| \bm d(\bm u) \|^2,
\end{equation}
where scaler $\bm\rho>0$ trades-off between the operational task space performance and power loss. Finally, the Newton/log-barrier method can proceed by fixing $\eta$ at a certain value and applying Newton's method to solve the following unconstrained problem
\begin{equation}
\min_{\bm u} \psi(\bm u, \eta) = \bm u^T \bm W' \bm u - \rho  \bm b^T \bm u -  \eta \sum_{i=1}^{r} \log c_i(\bm u)
\end{equation}
where $\bm b = 2 \bm B^T {\bm P} \bar{\bm M}^{-2} \bm \tau_c$ and $\bm W'$ is the new positive-definite weighing matrix
\begin{equation}
\bm W' = \bm W + \rho \bm B^T {\bm P} \bar{\bm M}^{-2}  {\bm P} \bm B.
\end{equation}

\subsection{Generalization of the constraints}
The method is general enough that additional equality or
inequality constraints can be included. For instance, in the case
of a walking robot with flat foot on the ground, the net moment on the foot is
the summation  of all positive normal force acting on the foot
times their distances from the pivot point. It is known that
different force distribution between the feet of the robot and
the ground through joint control is possible even if the feet are
constrained in the same fashion \cite{Chen-Huang-Yu-2010}. With proper location of the coordinate
frame  in the supporting convex hull for defining  the contact
force and moment, a ZMP-like criteria for dynamical stability of
the whole body  stipulates negative
 moment in $x$ and $y$ directions at the interface between the foot, i.e.,
$\lambda_{m_x}<0$ and $\lambda_{m_y}<0$. The latter inequalities
are tantamount to
\begin{equation} \label{eq:lamb_mxy}
\bm c'(\bm u)=-\bm A^+ \bm\Sigma_m  \bm S(\bm B \bm u + \bm\tau_g -\bm Q
\dot{\bm q})  \succeq  \bm 0
\end{equation}
where $\bm\Sigma_{m}$ is selection matrix corresponding to the
lagrangian multipliers of interest $\bm\lambda_m=[\lambda_{m_x}
\;\; \lambda_{m_y}]^T$, i.e., $\bm\lambda_{m} = \bm\Sigma_m \bm\lambda$. Therefore, inequalities \eqref{eq:lamb_mxy} and/or \eqref{eq:lamb_e} can be also included in the optimization formulation
\eqref{eq:quadratic_programming} to ensure dynamics stability of a
walking robot. Similarly, suppose a set of lagrangian multiplies $\bm\lambda_e =\bm\Sigma_{e} \bm\lambda$ are required to be regulated according to desired value $\bm\lambda_{e_d}$. Then this control objective can be achieved by including the  corresponding equality constrains in optimization formulation
\eqref{eq:quadratic_programming}
\begin{equation} \label{eq:lamb_e}
\bm\psi(\bm u)=-\bm A^+ \bm\Sigma_e  \bm S(\bm B \bm u + \bm\tau_g -\bm Q
\dot{\bm q}) -    \bm\lambda_{e_d} = \bm 0
\end{equation}

%------------------------------------------------------
\section{Conclusion}
%------------------------------------------------------
Energy-efficient controls of robots
subject to a variety of linear/nonlinear inequality constraints
such as Coulomb friction cones, unilateral contacts, actuator saturation limits has been presented. The optimal controller minimizes power dissipation in the joint motors
while constraining multiple contact forces to lie within their friction cones along trajectories of the operational space. As important aspect of the control scheme was that it did not require the  common linearization of friction cone constraints. Furthermore, the controller inherently can deal with snitching contact situation  as well as underactuation. Finally, the optimal problem problem to minimize  instantaneous power losses in the actuators has been transcribed as an quadratically constrained quadratic programming (QCQP) for the operational space control of the robot while satisfying all linear and nonlinear constraints  in order to avoid slipping at contact surfaces, to maintain the unilateral contacts, and to differ from actuator saturation. It was shown that the optimization formulation naturally allowed trading-off the operational space task when  it was not possible to enforce it along with all physical limitations so that the perturbation to the tracking performance was minimized. Experimental results obtained from a robot interacting with a contact friction surface comparatively demonstrated significant
reduction in power losses. In spite of their versatility in traversing many different terrains, multi-legged robots  require increased power consumption compared to their wheeled robot counterpart. Therefore, the optimal QCQP-based controller is proposed for future implementation on  multi-legged robots  to demonstrate power saving.

\appendix

%=====================================================
\subsection{Time-derivative of projection matrix} \label{appx_skew}
%=====================================================
The Tikhonov regularization theorem \cite{Golub-VanLoan-1996} describes the pseudo-inverse as  the following limit
\begin{equation} \label{eq:Tikhonov}
\bm A^+ = \lim_{\epsilon \rightarrow 0} \bm A^T(\bm A \bm A^T + \epsilon \bm I)^{-1}
\end{equation}
By differentiation of  the above expression, one can verify that   the time-derivative of the  pseudo-inverse can written in the following form
\begin{equation} \label{eq:dotA+}
\frac{d}{dt} \bm A^+ = -\bm A^+ \dot{\bm A} \bm A^{+} + \lim_{\epsilon \rightarrow 0} \bm P \dot{\bm A}^T (\bm A \bm A^T + \epsilon \bm I)^{-1}
\end{equation}
On the other hand, using \eqref{eq:dotA+} in the time-derivative of the expression of the projection matrix $\bm P= \bm I - \bm A^+ \bm A$ yields
\begin{equation} \label{eq:dotP-A+}
\begin{split}
\dot{\bm P} &= - \frac{d}{dt}\bm A^+ \bm A - \bm A^+ \dot{\bm A} \\ \notag
& =  \bm A^+ \dot{\bm A} \bm A^{+} \bm A + \lim_{\epsilon \rightarrow 0} \bm P \dot{\bm A}^T (\bm A \bm A^T + \epsilon \bm I)^{-1} \bm A - \bm A^+ \dot{\bm A} \\
& =\bm A^+ \dot{\bm A} (\bm I - \bm P) + \bm P \dot{\bm A}^T \bm A^{+T} - \bm A^+ \dot{\bm A} \\
&=  \bm L + \bm L^T
\end{split}
\end{equation}
where $\bm L=- \bm A^+ \dot {\bm A} \bm P$. Note that identity $\bm P \bm A^+ = \bm A^{+T} \bm P = \bm 0$  implies that $\bm L^T \bm P = \bm 0$ and hence one can conclude $\ddot{\bm q}_{\perp}= \dot{\bm P} \dot{\bm q}= \bm L \dot{\bm q} = \bm\Omega \dot{\bm q}$.

%=====================================================
\subsection{Dynamics formulation}
\label{appx:formulation}
%=====================================================
Substituting the generalized acceleration from $\ddot {\bm
q}= \bm P \ddot{\bm  q} + \dot{\bm  P} \dot{\bm q}$ into  \eqref{eq:projected_dynamics} gives
\begin{equation} \label{eq:PMPddq}
\bm P \bm M \bm P \ddot{\bm  q} + \big(\bm P \bm C \bm P + \bm P
\bm M \dot{\bm  P} \big)  \dot{\bm q}= \bm P \bm B \bm u + \bm P \bm\tau_g.
\end{equation}
Now pre-multiplying  equation \eqref{eq:Omega_dq} by positive scaler $\nu$ and then adding both sides of the latter equation with those of \eqref{eq:PMPddq} and using identity $\bm\Gamma^T \bm P = \bm 0$, one can obtain  \eqref{eq:constraint_system}.

The following analysis will show that the fundamental
properties of constrained inertia matrix derived (as stated in
Property~\ref{eq:prop_constraint_mass}) are similar to those of
general inertia matrix of the system. Consider non-zero vector
$\bm a \in \mathbb{R}^n$ and its orthogonal decomposition components $\bm a_{\parallel}=\bm P \bm a$ and $\bm a_{\perp}=(\bm I - \bm P) \bm a$. Then, one can say
\begin{equation} \label{eq:twoMass}
\bm a^T \bar{\bm M} \bm a  =
\bm a^T_{\parallel} \bm M \bm a_{\parallel} + \nu \| \bm a_{\perp}
\|^2
>0,
\end{equation}
Notice that both terms $\bm a^T_{\parallel} \bm M \bm a_{\parallel}>0$ and $\| \bm a_{\perp}
\|^2>0$ are positive semi-definite. Moreover, for a given non-zero vector $\bm a$ if $\bm a_{\perp}= \bm 0$ then $\bm a^T_{\parallel} \neq 0$ and vice versa. This means that the summation of the two orthogonal terms must be positive definite and so must be the constraint inertia matrix $\bar{\bm M}$.

Furthermore, subtracting the expression of $2\bar{\bm C}$ given in \eqref{eq:Cbar2} from the the time-derivative of the constraint inertia matrix  obtain from   \eqref{eq:Mbar} leads to  the following matrix equations
\begin{align} \notag
\dot{\bar{\bm M}} - 2 \bar{\bm C}  =& \dot{\bm P} \bm M \bm P + \bm P \bm M \dot{\bm P} + \bm P \dot{\bm M} \bm P + \nu \dot{\bm P} \\ \notag
& - 2 \bm P \bm C \bm P -2 \bm P \bm M(\bm\Gamma +  \bm\Gamma^T) + 2 \nu \bm\Gamma \\ \notag
 = & \bm P[\dot{\bm  M} - 2\bm C] \bm P + [\bm\Gamma^T
\bm M \bm P - (\bm\Gamma^T \bm M \bm P)^T] \\ \label{eq:M-2C}
&+[\bm\Gamma \bm M \bm P - (\bm\Gamma \bm M \bm
P)^T ] + \nu[\bm\Gamma  - \bm\Gamma^T]
\end{align}
One can verify that all matrix terms  in the brackets of \eqref{eq:M-2C} are skew-symmetric and hence so is the matrix $\dot{\bar{\bm M}} - 2 \bar{\bm C}$.

In the remainder of this appendix, we will derive an upper bound for the constraint inertia matrix. Let us consider the characteristic equation of the matrix $\bar{\bm M}$ as
\begin{equation} \notag
\bar{\bm M} \bm x - \lambda \bm x = \big( \bm P \bm M \bm P + \nu (\bm I - \bm P) \big) \bm x - \lambda \bm x = \bm 0
\end{equation}
Clearly $\lambda=\nu$ is the eigenvalue for all orthogonal eigenvectors which span ${\cal N}^{\perp}$ because $\lambda=\nu$ means $(\bm P \bm M \bm P - \bm P) \bm x =\bm 0 \quad \forall \bm x \in {\cal N}^{\perp}$. The remaining set of  orthogonal eigenvectors  must lie in ${\cal N}$ that are  corresponding to the non-zero eigenvalues of $\bm P \bm M \bm P$
\begin{equation} \notag
\bm P \bm M \bm P \bm x - \lambda \bm x =\bm 0 \qquad  \lambda \neq 0 \quad \forall \bm x \in {\cal N}
\end{equation}
Therefore, the set of all eigenvalues of the p.d. matrix $\bar{\bm M}$ is  the union of the above sets corresponding to the eigenvectors in ${\cal N}$ and ${\cal N}^{\perp}$, i.e.,
\begin{equation} \notag
\lambda(\bar{\bm M})=: \big\{\underbrace{\nu, \cdots, \nu}_{r}, \;  \underbrace{ \lambda_{\stackrel{\rm min}{\neq0}} (\bm P \bm M \bm P), \cdots, \lambda_{\rm max}(\bm P \bm M \bm P)}_{n-r} \big\}
\end{equation}
where $\{ \lambda_{\stackrel{\rm min}{\neq0}}(\bm P \bm M \bm P), \cdots, \lambda_{\rm max}(\bm P \bm M \bm P) \}$ are all non-zero eigenvalues of $\bm P \bm M \bm P$. Thus
\begin{equation} \notag
\| \bar{\bm M} \| = \lambda(\bar{\bm M}) = \max(\nu, \| \bm P \bm M \bm P \| ) \leq \max(\nu, \| \bm M  \| ),
\end{equation}
which implies bounded $\bar{\bm M}$ because both $\bm M$ and $\nu$ are bounded quantities.

%=====================================================
\subsection{Constraint forces}
\label{appx:lagrangian}
%=====================================================
 Let us obtain the generalized acceleration from \eqref{eq_EqMotion2} using the inverse of the constraint inertia matrix and then substitute the  solution   into \eqref{eq:Mddq}. Then,
\begin{align}\notag
\bm A^ T \bm\lambda =& \bm M \ddot{\bm q} + \bm C \dot{\bm q} -(\bm u + \bm\tau_g) \\ \notag
 =& \bm M \bar{\bm M}^{-1} \bm P (\bm B \bm u - \bm\tau_g - \bm C \dot{\bm q}) + \bm M \bar{\bm M}^{-1}(\nu \bm I - \bm P \bm M) \bm\Gamma \dot{\bm q} \\ \notag
=&\bm M \bar{\bm M}^{-1} \bm P (\bm B \bm u - \bm\tau_g - \bm C \dot{\bm q}) + (\bm M - \bm M \bar{\bm M}^{-1} \bm M) \bm\Gamma \dot{\bm q} \\ \label{eq:ATlamb}
=& -\bm S \Big(\bm B \bm u + \bm\tau_g - ( \bm M \bm\Omega + \bm C) \dot{\bm q}  \Big)
\end{align}
where the non-symmetric matrix $\bm S$ has been perviously defined in \eqref{eq:S_definition}.
Note that \eqref{eq:ATlamb} is obtained by using identities  $\bm\Gamma\dot{\bm q} =\bm\Omega \dot{\bm q}$ and $\nu \bar{\bm M}^{-1} \bm\Gamma \dot{\bm q} = \bm\Gamma \dot{\bm q}$, which in turn the is inferred from
$({\bm I - \bm P}) \bm\Gamma = \bm\Gamma$, and $\nu \bar{\bm M}^{-1}(\bm I - \bm P) = (\bm I - \bm P)$. From definition, we have
\begin{equation} \notag
\bm P \bar{\bm M} = \bar{\bm M} \bm P
\end{equation}
meaning that the projection matrix and the constraint inertia matrix commute. It other words,  $\bar{\bm
M} \bm P \bar{\bm M}^{-1} =\bm P$ or equivalently  $\bm P \bm M \bm P\bar{\bm M}^{-1} \bm P= \bm
P$, which in turn is used in the following derivation
\begin{equation} \notag
\bm S^2 = \bm I -2 \bm M \bar{\bm M}^{-1} \bm P +\bm M \bar {\bm
M}^{-1} \underbrace{\bm P \bm M \bm P \bar{\bm M}^{-1} \bm P}_{\bm  P} = \bm S.
\end{equation}

%=====================================================
\subsection{Properties of $\bm\Lambda$} \label{appx:Lambda}
%=====================================================
It can be inferred from \eqref{eq_dtheta} that $\bm\Lambda \bm P = \bm\Lambda$ whose transpose readily gives \eqref{eq:PLambdaT}.

Because independent set $\bm x(\bm q) \in \mathbb{R}^l$ is selected for the operational space control, the Jacobian matrix $\bm\Lambda$ has linearly independent rows and hence its pseudo-inverse has the form $\bm\Lambda^+=\bm\Lambda^T(\bm\Lambda \bm\Lambda^T)^{-1}$ meaning that
\begin{equation} \notag
{\cal R}(\bm\Lambda^+) = {\cal R}(\bm\Lambda^T) = {\cal N}^{\perp}(\bm A),
\end{equation}
and hence \eqref{eq:PLambda+} is in order.

Pre-multiplying both sides of \eqref{eq:ddx} by $\bm\Lambda^+$, we get
\begin{equation} \label{eq:Lambdaddq}
\bm\Lambda^+ \ddot{\bm x} = \bm\Lambda^+ \dot{\bm\Lambda} \dot{\bm q} + \bm P \ddot{\bm q}
\end{equation}
in which we obtain the identity $\bm\Lambda^+\bm\Lambda = \bm P$ from  ${\cal N}^{\perp}(\bm\Lambda) = {\cal N}(\bm A)$ that, in turn, is inferred from ${\cal R}(\bm\Lambda^T) = {\cal N}(\bm A)$. Finally, adding both sides of equations \eqref{eq:Omega_dq} and  \eqref{eq:Lambdaddq}, the latter equation can be rearranged in the form \eqref{eq:Lambda+ddx}, i.e.,
\begin{equation}
\bm\Lambda^+ \ddot{\bm x} - (\bm\Lambda^+ \dot{\bm\Lambda}  -\bm\Omega) \dot{\bm q} = \ddot{\bm q},
\end{equation}

%=====================================================
\subsection{Interior point method} \label{appx:interior_point}
%=====================================================
Suppose $r=2(k+p)$ be the total number of inequalities. It has
shown that $r \eta$ is the duality gap, and therefore the
central path $\bm u^*(t)$ converges to the optimal solution as
$\eta \rightarrow 0$. In the interior point methods, Newton's
method  is used to find a sequence of solution $\bm u^*(\eta)$
which  follows the central path by increasing the values of $\eta$.
This method follows the following steps: Starts with a large value
$\eta>0$, decrement gain $\kappa<1$, and a given strictly feasible
$\bm u$ as the initial value. For a give precision $\epsilon$,
iterate the following steps
\begin{enumerate}
\item Set the initial value as $\bm u=\bm B^T
{\bm\tau}_c$.
 \item For the given value of $\eta$, solve the barrier problem  \eqref{barrier_problem}
using Newton's method to find $\bm u^*(\eta)$.
\label{step_barrier}
 \item Update $\bm u = \bm u^*(\eta)$
\item Stop if $r \eta \leq \epsilon$
\item Decrease the
barrier variable $\eta:=\kappa \eta$, and go to step \ref{step_barrier}.
\end{enumerate}

%-------------------------------------------------------------
\bibliographystyle{IEEEtran}
%\bibliography{references}
%-------------------------------------------------------------

\end{document}